\documentclass[twocolumn]{aastex62}
\usepackage{graphics,epsf}
\usepackage{amsmath}                
\usepackage{amsfonts}               
\usepackage{amssymb}                
\usepackage{epsfig}                 
\usepackage{graphicx}
\usepackage{float}
\usepackage{color}
\usepackage[para,online,flushleft]{threeparttable}

\newcommand{\cm}{{~\rm cm}}
\newcommand{\km}{{~\rm km}}
\newcommand{\s}{{~\rm s}}

\newcommand{\g}{{~\rm g}}

\newcommand{\K}{{~\rm K}}
\newcommand{\erg}{{~\rm erg}}
\newcommand{\yr}{{~\rm yr}}

\newcommand{\AU}{{~\rm AU}}

\begin{document}

\title{Three-dimensional simulations of the jet feedback mechanism in common envelope jets supernovae}

\author{Shlomi Hillel}
\affiliation{Department of Physics, Technion, Haifa, 3200003, Israel; \\ ronsr@technion.ac.il, shlomi.hillel@gmail.com, soker@physics.technion.ac.il}

\author{Ron Schreier}
\affiliation{Department of Physics, Technion, Haifa, 3200003, Israel; \\ ronsr@technion.ac.il, shlomi.hillel@gmail.com, soker@physics.technion.ac.il}

\author[0000-0003-0375-8987]{Noam Soker}
\affiliation{Department of Physics, Technion, Haifa, 3200003, Israel; \\ ronsr@technion.ac.il, shlomi.hillel@gmail.com, soker@physics.technion.ac.il}

\begin{abstract}
We conduct three-dimensional hydrodynamical simulations of common envelope jets supernova (CEJSN) events where we assume that a neutron star (NS) launches jets as it orbits inside the outer zones of a red supergiant (RSG) envelope, and find the negative jet feedback coefficient to be $\simeq 0.1-0.2$. This coefficient is the factor by which the jets reduce the mass accretion rate onto the NS as they remove mass from the envelope and inflate bubbles (cocoons). Our results suggest that in most CEJSN events the NS-RSG binary system experiences the grazing envelope evolution (GEE) before it enters a full common envelope evolution (CEE). We also find that the jets induce upward and downward flows in the RSG envelope. These flows together with the strong convection of RSG stars might imply that energy transport by convection in CEJSNe is very important. Because of limited numerical resources we do not include in the simulations the gravity of the NS, nor the accretion process, nor the jets launching process, and nor the gravity of the deformed envelope. Future numerical simulations of CEE with a NS/BH companion should include the accretion process onto the NS (and vary the jets’ power accordingly), the full gravitational interaction of the NS with the RSG, and energy transport by the strong convection. 
\end{abstract}


\section{INTRODUCTION}
\label{sec:intro}

In the common envelope jet supernova (CEJSN) scenario a neutron star (NS) or a black hole (BH), hereafter NS/BH, enters the envelope of a giant star, accretes mass through an accretion disk that launches jets, and spirals-in to the core of the giant star (e.g., \citealt{SokerGilkis2018, Gilkisetal2019, Sokeretal2019CEJSN, GrichenerSoker2019a, LopezCamaraetal2019, LopezCamaraetal2020MN, Soker2021, Metzger2022}). The NS/BH spirals-in in this common envelope evolution (CEE), enters the core and further accretes mass from the core (for NS/BH merger with the core without explicitly including jets see, e.g., \citealt{FryerWoosley1998, ZhangFryer2001, BarkovKomissarov2011, Thoneetal2011, Chevalier2012, Schroderetal2020}). Eventually the NS/BH tidally destroys the core such that the core material forms a massive accretion disk around the NS/BH. The interaction of the energetic jets that the NS/BH launches, in particular after it reaches the core, with the envelope leads to a bright transient event that can mimic a supernova. This is a CEJSN event. If the NS/BH accretes mass from the tenuous envelope to power a bright transient event but it does not spiral-in all the way to the core, the event is termed a CEJSN impostor (e.g., \citealt{Gilkisetal2019, Soker2022FBOT}.

There are two key processes that enable the CEJSN (or the CEJSN impostor) scenario. The first one is neutrino cooling that allows a very high mass accretion rate when the mass accretion rate is $\dot M_{\rm acc} \ga 10^{-3} M_\odot \yr^{-1}$ \citep{HouckChevalier1991, Chevalier1993, Chevalier2012}. The second process is that the NS/BH accretes mass through an accretion disc (e.g.,  \citealt{ArmitageLivio2000, Papishetal2015, SokerGilkis2018}). 
Different studies reach different values of the mass accretion rate (mainly whether it is close to the Bondi-Hoyle-Lyttleton [BHL] value of $\dot M_{\rm BHL}$ or whether it is much smaller), and on whether the accreted mass forms an accretion disk around the compact object (e.g., \citealt{RasioShapiro1991, Fryeretal1996, Lombardietal2006, RickerTaam2008, Shiberetal2016, MacLeodRamirezRuiz2015a, MacLeodRamirezRuiz2015b,  MacLeodetal2017}).  
We will follow recent studies that show the formation of an accretion disk around the compact companion in a CEE (e.g., \citealt{Chamandyetal2018}), in particular for a NS companion (\citealt{LopezCamaraetal2020MN}).  

The case of NS/BH accretion is very different from cases of a main sequence star that spirals-in inside the envelope of a giant star. In particular relevant to our study are these two differences. (1) Main sequence stars have a hard time forming an accretion disk because of their relatively large radius (e.g., \citealt{LopezCamaraetal2021} for  recent simulations). NS/BH are much smaller and do not encounter this problem. (2) Main sequence stars have difficulties in launching jets deep inside the envelope as the jets are choked (e.g., \citealt{LopezCamaraetal2021}). The much deeper potential well of NS/BHs results in much more energetic jets that can expand to some distance in the envelope, even if they do not break out from the envelope (e.g., \citealt{GrichenerSoker2021}). 
\cite{LopezCamaraetal2019} and \cite{LopezCamaraetal2020MN} find the NS companion in their CEE simulations to accrete at a rate of $\dot M_{\rm acc} \approx 0.1 - 0.2 \dot M_{\rm BHL}$. We will consider even lower mass accretion rates in our study. We also note that the equation of state of the NS influences the outcome of accretion (e.g., \citealt{Holgadoetal2021}). 

The CEJSN scenario might account for some puzzling transient events and processes in astrophysics (e.g., \citealt{Sokeretal2019CEJSN}). \cite{Thoneetal2011} suggested that a merger of a NS and a helium star, which we would classify as a CEJSN event, can account for the unusual Gamma-ray burst GRB~101225A. 
\cite{SokerGilkis2018} proposed that the enigmatic supernova iPTF14hls \citep{Arcavietal2017} was a CEJSN event. The CEJSN might account also for SN~2020faa that is a similar supernova to iPTF14hls (see \citealt{Yangetal2020} for the observations of SN~2020faa). 
\cite{Sokeretal2019CEJSN} and \cite{Metzger2022} proposed the CEJSN scenario, and \cite{Soker2022FBOT} the CEJSN impostor scenario,  for the fast-rising blue optical transient AT2018cow. The CEJSN might be one of the sites for the r-process nucleosynthesis \citep{Papishetal2015, GrichenerSoker2019a, GrichenerSoker2019b, Gricheneretal2022}. As well, CEJSNe with BHs companions that launch the jets might be one of the sources of high-energy neutrinos \citep{GrichenerSoker2021}. 
Based on these earlier studies \cite{Dongetal2021} suggested recently a CEJSN scenario for the luminous radio transient, VT~J121001+495647.  

Earlier studies found that jets in CEE with main sequence companions can remove a substantial amount of envelope mass when the main sequence star is in the outer parts of the envelope (e.g., \citealt{Shiberetal2019}). This can increase the common envelope efficiency above unity, i.e., $\alpha_{\rm CE} > 1$. Naturally NS/BH companions that launch jets can also bring the CEE efficiency to $\alpha_{\rm CE} > 1$, as some scenarios require for the CEE with a NS or a BH companion (e.g. \citealt{Fragosetal2019, Zevinetal2020, Garciaetal2021}). In any case, our view is that jets that the NS/BH launches play a significant role in the CEE, 
and CEE simulations that involve a NS companion (e.g., \citealt{LawSmithetal2021}) should eventually include jets to give a more accurate picture of the CEE.

Our main goal is to estimate the value of \textit{negative jet feedback coefficient} in our simulations (section \ref{sec:Feedback}). This coefficient is the further reduction in the mass accretion rate because the jets inflate the envelope and eject mass, such that the density in the NS vicinity decreases. We do not calculate the full negative feedback cycle, but rather we follow the one-dimensional (1D) simulations of \cite{Gricheneretal2021} to estimate that coefficient, but now with 3D simulations. 

We set the NS source of the jets to orbit inside the envelope of a red supergiant (RSG) star (section \ref{sec:Numerical}) and simulate several different cases (section \ref{sec:Cases}). We do not include either the self-gravity of the envelope nor that of the NS. For detailed CEE simulations of a NS inside a RSG envelope that include these effects, but that do not include jets, see the recent studies by \cite{Lauetal2022} and \cite{Morenoetal2022}. We also do not follow the accretion process. We preset the jets' power and the orbital motion. As well, we set the jets to mix with the envelope mass in two cones about the equatorial plane, as we cannot resolve the launching zone of the jets \citep{Schreieretal2021}.   These omissions allow us to follow the system for years, unlike simulations that include the accretion process onto the NS/BH and are therefore limited to evolution time of several days or less (e.g., \citealt{MorenoMendezetal2017, LopezCamaraetal2019, LopezCamaraetal2020MN}).    
To reach our goal we set the numerical scheme in a similar manner to earlier studies of jet-induced outflow in CEE (e.g., \citealt{ShiberSoker2018, Schreieretal2019}), but in the present study we include jets from a NS rather than from a main sequence star. 
We follow the flow structure in sections \ref{sec:flow} and in section \ref{sec:Feedback} we study the negative feedback process. 
 We discuss the uncertainties of our simulations in section \ref{sec:Uncertainties},  and summarise our main results in section \ref{sec:Summary}. 
     
\section{Numerical set up}
\label{sec:Numerical}

\subsection{Basic simplifying assumptions}
\label{subsec:assumptions}

To facilitate long simulations, i.e., of several orbital periods, and to reach our goal of determining the negative jet feedback coefficient, we make the following basic assumptions. 

\subsubsection{A spherical star}
\label{subsubsec:spherical}
We use the stellar evolution code \texttt{MESA} \citep{Paxtonetal2011, Paxtonetal2013, Paxtonetal2015, Paxtonetal2018, Paxtonetal2019} to evolve a zero-age-main-sequence star of 
$M_{\rm 1,ZAMS}=15 M_\odot$ for $1.1\times10^6 \yr$, when it becomes a RSG star with a radius of $R_{RSG}=881\,R_{\odot}$, 
a mass of $M_1=  12.5 M_\odot$, 
metallicity of $Z=0.02$, and an effective temperature of $T_{\rm eff}= 3160K$. 
We take this RSG stellar model and install it at the center of our computational grid of the three-dimensional (3D) hydrodynamical code {\sc flash} \citep{Fryxelletal2000}. 
The density in the outer boundary of the initial RSG model is $\rho(R_{RSG}) = 2.1 \times 10^{-9} \g \cm^{-3}$. The composition of the 3D star is pure hydrogen, and the code
assumes full ionisation. 
For numerical reasons we fill the initial numerical grid volume outside the star with gas of density $\rho_{\rm grid,0} = 2.1 \times 10^{-13} \g \cm^{-3}$ and of temperature $T_{\rm grid,0}= 1100 \K$.  We use this non-rotating spherical stellar model and neglect the process by which the NS spins-up the envelope before it enters the envelope.
 
\begin{figure} 
\centering
\includegraphics[width=0.46\textwidth]{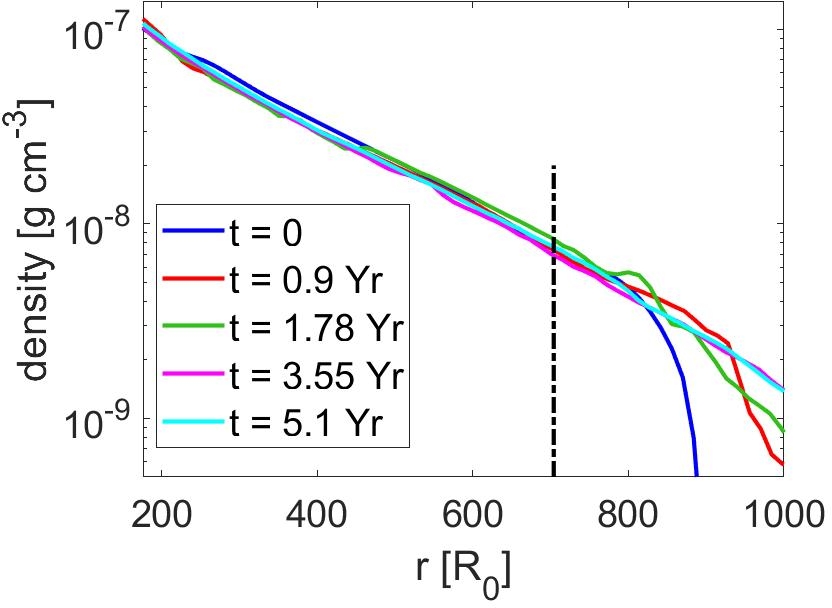}
\caption{ The density profiles at five times in the simulation without jets. The density profiles are along a particular radial line from the outer boundary of the inert inner core to $r=1000 R_\odot$. The vertical dashed-dotted line is the location of the NS orbit. We are interested in the density at that radius only.  }
\label{fig:testcase} 
\end{figure}

We are interested in the density of the star at the orbit of the jets' source (the NS). As we show in Fig. \ref{fig:testcase} and also later in section \ref{sec:Feedback}, in a simulation without jets the value of this density fluctuates in the first three years, but then settled to a practically constant value. For that, we calculate the densities as a results of jets during the time after relaxation. We are interested only in the stellar density at and near the orbit of the NS, which we mark by a vertical dashed-dotted line on Fig. \ref{fig:testcase}. We learn from Fig. \ref{fig:testcase} that although the stellar model does not maintain its density profile in the outer regions, near the orbit of the NS the stellar model is adequately stable for our goals (see also section \ref{sec:Feedback}). 

\subsubsection{A circular orbit}
\label{subsubsec:circular}
We simulate bipolar outflows that result from jets inside the envelope of a RSG star. We attribute these jets to a NS that orbits inside the envelope of the RSG star and accretes mass. We do not calculate the orbital in-spiral, and assume a circular orbit with a constant radius of $a= 700 R_\odot=0.79 R_{RSG}$. We justify this simplifying assumption as our goals are to explore the flow structure in the envelope due solely to jets and to find the negative jet feedback coefficient. For the same reason we do not include the orbital energy that the NS releases as it spirals-in. 
    
\subsubsection{An inert inner core}
\label{subsubsec:inert}
To save expensive computational time we replace the inner $20\%$ of the stellar radius, $R_{\rm in} = 176\,R_{\odot}$, with an inert ball having constant shape, density, pressure, and temperature. We do include the gravitational potential of this inner inert ball of mass $5.64  M_\odot$. 

\subsubsection{A constant gravitational potential}
\label{subsubsec:potential}
The gravitational field in our simulation is constant in time and equals to that of the initial RSG star. Namely, we neglect both the gravity of the NS and of the deformed envelope. Again, the goal of this study is to explore the role of jets alone. 

\subsubsection{The jet-envelope interaction in the NS vicinity}
\label{subsubsec:interaction}
  
The NS would be expected to launch jets at $\approx 10^5 \km \s^{-1}$. In addition, the NS is moving inside the envelope and we would like to resolve the entire RSG and its vicinity. Our computer resources do not allow us to resolve the region near the NS, and therefore we cannot accurately follow the early interaction of the jets with the envelope. To circumvent these difficulties we follow  \cite{Schreieretal2021} and assume the following. 

The NS launches very fast jets with a combined power of $\dot E_{\rm 2j}$. We take the jet-envelope interaction to occur inside two opposite lobes, each with a conical shape on opposite sides of the equatorial plane, with a combined volume of $V_{\rm 2j}$. The length of each lobe is $L_{\rm L} = 7\times 10^{12} \cm$ and its half opening angle is $\alpha_{\rm L}= 30^\circ$. At the beginning of the time step the gas inside these cones has a total mass of $M_{\rm 2c,i}$, a total kinetic energy of $E_{\rm 2kc,i}$, and a total thermal energy of $E_{\rm 2tc,i}$. We neglect the change in gravitational energy (as we neglect the self-gravity of the envelope) and the mass that the NS accretes from its vicinity and the mass that the jets add to the cones, as these are small in the specific simulations that we conduct.
Namely, the jets' injection scheme does not change the mass in the injection zone (the two cones), but rather only the energy and momentum there.  

This scheme is based on the finding that the jets that the NS/BH launches in CEJSNe do not penetrate through the envelope when the NS/BH is not too close to the RSG surface (e.g., \citealt{Papishetal2015, GrichenerSoker2021}). The reason is that the NS/BH orbits inside the RSG envelope. Therefore, the NS/BH supplies new jet's material on an ever changing line. It turns out that the time for a jet to breakout from the RSG surface is much longer than the time the jet receives a new supply of momentum (see these papers). Therefore, the jets are choked inside the envelope.  

The very fast jets that the NS launches carry a relatively low momentum. We assume that the momentum in the lobes is mainly due to the pressure gradient that the shocked jets build near the center. We further take the jet-envelope interaction near the NS to be like an explosion, and set the kinetic energy and the thermal energy after interaction to be equal to each other $E_{\rm 2kc}=E_{\rm 2tc}$.
The velocity of the gas in the cones is radial of value $v_{\rm c,r}$.

According to these assumptions then, the new properties of the gas inside the cones after jet-envelope interaction within a time step $\Delta t$ are 
\begin{eqnarray}
\begin{aligned}
& M_{\rm 2c}=M_{\rm 2c,i}  
\\ &
E_{\rm 2c}=E_{\rm 2kc,i}  + E_{\rm 2tc,i} + \dot E_{\rm 2j} \Delta t
\\ &
E_{\rm 2kc}=0.5E_{\rm 2c}; \qquad E_{\rm 2tc}=0.5E_{\rm 2c};
\\ &
v_{\rm c,r} = \sqrt{2 E_{\rm 2kc}/ M_{\rm 2c} },
\end{aligned}
\label{eq:Cones}
\end{eqnarray}
where the subscript `i' refers to the beginning of the time step. 

\subsection{The computational scheme}
\label{subsec:Grid}

\subsubsection{The grid}
\label{subsubsec:grid}
The computational numerical grid is a cube with a side of 
$L_{\rm G}$. The adaptive-mesh-refinement (AMR) of the numerical code FLASH divides to smaller cell sizes at zones of high gradients. We list the values of $L_{\rm G}$ and the sizes of the grid-cells for the different simulations in section \ref{sec:Cases}. 
In the entire computational grid the equation of state of the gas is that of 
an ideal gas with an adiabatic index of $\gamma=5/3$ plus radiation pressure. 
The centre of the RSG does not change and it is at $(x_1,y_1,z_1)=(0,0,0)$. 

The refinement criterion is the default error estimator in FLASH (modified L\"ohner estimator with default parameters) on the $z$-component of the velocity. We take the $z$-component of the velocity because we launch fast jets along the $\pm z$-directions. The highest resolution simulation, RUN39, starts with about three million cells, and as the envelope expands the number of cells increases to about 192 million cells through three levels of refinement.
The lowest simulation runs (RUN41L and RUN42L) have 
about $3 \times 10^5$ cells with two levels of refinement.

The ratios of the pressure scale-height at the NS orbit, $H_{p} = 0.14r=100 R_\odot$, to the smallest grid-cell size are in the range of 7.1 in the high resolution simulation to  1.8 in the lowest resolution simulations.   

\subsubsection{Accretion rates}
\label{subsubsec:Accretion}

We set the orbital motion of the source of the jets, the NS, to have a Keplerian circular orbit with a radius of $a= 700 R_\odot$ and an orbital period of $1.77 \yr$.  
The density of the unperturbed envelope at this radius is 
$\rho_0(700)= 7.9 \times 10^{-9} \g \cm^{-3}$ and 
the mass of the giant star inner to this orbit is  $M_1(700)= 11.3 M_\odot$.
The relative velocities of the NS and the envelope, neglecting envelope rotation, is $v_{\rm NS}(700)= [G (M_{\rm NS} + M_1(a))/a]^{1/2} = 59 \km \s^{-1}$. 
The BHL accretion rate from the envelope  
\begin{equation}
\dot M_{\rm BHL} = \pi \rho(a) v_{\rm rel} (a)
\left[ \frac{2 G M_{\rm NS}}{v^2_{\rm NS}(a)} \right]^2 ,
\label{eq:MBHL}
\end{equation}
in the unperturbed envelope for $a=700 R_\odot$ is  
\begin{equation}
\dot M_{\rm BHL,0} (700) = 0.27 M_\odot \yr^{-1},
\label{eq:MBHL0}
\end{equation}
where we take for the NS mass $M_{\rm NS} =1.4 M_\odot$ and subscript `$0$' refers to the BHL accretion rate in the unperturbed envelope. For the relative velocity of the NS inside the envelope we take $v_{\rm NS}$, although the velocity would be somewhat lower if envelope rotation were included. On the other hand, inclusion of the sound speed in the envelope reduces the accretion rate. 
 
For the above mass accretion rate the gravitational accretion power is $\dot E_{\rm acc, BHL,0} (700)=  2.6 \times 10^{45}  \erg \s^{-1}$ 
for a NS with a radius of $12 \km$.
We expect the power of the jets that the accretion disk launches to be 
\begin{equation}
\dot E_{\rm 2j} = \zeta \dot E_{\rm acc, BHL,0} = \zeta \frac {G M_{\rm NS} \dot M_{\rm BHL,0}}{R_{\rm NS}}  ,
\label{eq:E2jzeta}
\end{equation}
with $\zeta \approx 10^{-3}-10^{-2}$ \citep{Gricheneretal2021} because of the negative jet feedback mechanism (for a review see \citealt{Soker2016Rev}). Nonetheless, because of numerical limitations we simulate cases with much lower jets' power and up to $\zeta =5 \times 10^{-4}$.
We return to the negative jet feedback mechanism in section \ref{sec:Feedback}.

\section{Simulated cases}
\label{sec:Cases}

We summarise the different simulations we conduct in Table \ref{Table:cases}.
We list, from left to right, the name of the simulation, the combined power of the two jets, the ratio of this power to the unperturbed BHL accretion power $\zeta$ (equation \ref{eq:E2jzeta}), the full size of the numerical grid $L_{\rm G}$, and the cell sizes. The presence of more than one size of cells is due to the AMR grid, with smaller cells in regions of high gradients. These sizes represent the resolutions of the numerical code.  The first five columns are the input parameters of the simulations.  
In the sixth column we list the average factor by which the jets reduce the density in the envelope as we explain in section \ref{sec:Feedback}. We then list the figures in which we present the results from the respective simulations.   
The suffix `39' in the name of RUN39 stands for the exponent of the jets' power, and so forth. The Letter `L' stands for the lowest resolution we use. 
\begin{table*}[htbp]
\centering
\begin{tabular}{|c|c|c|c|c|c|c|c|}
\hline
 Run  & $\dot E_{\rm 2j}$ & $\zeta $    & $L_{\rm G}$   & 
 Cell sizes  & $q_\rho$ & Figures & comments \\ 
  &   Jets' energy & Equation (\ref{eq:E2jzeta}) & Grid Size & Resolutions & Density decrease & & \\  
   & $[\erg \s^{-1}]$  &  & $[R_\odot]$ & $[R_\odot]$ &          &         &           \\
 \hline 
RUN39  & $3.16 \times 10^{39}$ &  $10^{-6}$  & 
7184 & 56, 28, 14 & 0.8684 &   \ref{fig:AverageD}, \ref{fig:jetspower} &  \\  
 \hline
RUN40  & $3.16 \times 10^{40}$ &  $10^{-5}$  & 
7184 & 56, 28     & 0.5248 &  \ref{fig:RUN40profiles}, \ref{fig:AverageD},\ref{fig:jetspower}  & \\  
 \hline 
RUN41 & $3.16 \times 10^{41}$ &  $10^{-4}$  & 
3592 & 74.8, 37.4 & 0.4240 & \ref{fig:AverageD},\ref{fig:jetspower}  & Medium resolution\\  
 \hline 
RUN41L & $3.16 \times 10^{41}$ & $10^{-4}$ & 
3592 & 112, 56    & 0.3011 & \ref{fig:RUN41vs42} - \ref{fig:jetspower} & \\  
 \hline  
RUN42L & $1.58 \times 10^{42}$ & $5\times 10^{-4}$ & 
3592 & 112, 56    &  0.3832  & \ref{fig:RUN41vs42} - \ref{fig:jetspower}  &  A 19 weeks simulation \\  
 \hline 
 \end{tabular} 
\caption{The cases we simulate. In the second column we list the combined power of the two jets, and in the third column the value of $\zeta$ as we define in equation (\ref{eq:E2jzeta}). We make use of the density reduction factor by jets $q_{\rho}$ in section \ref{sec:Feedback}. For other parameters see the text.  RUN42L took us 19 weeks to simulate, and it is on the upper end of the capabilities of our numerical resources.
}
\label{Table:cases}
\end{table*}

We did not simulate more powerful jets because of our limited computational resources. In reality, we expect the jets to be more powerful than the power we use in simulations RUN42L (section \ref{sec:Feedback}). However, with increasing jets' power the flow velocities are higher and computational time becomes very long. In the cases of the high power simulations RUN41, RUN41L, and RUN42L we had to reduce the resolution. 
We cannot afford more powerful jets than in simulation RUN42L as then the much lower required numerical resolution would make the results unreliable. 
    
\section{The flow structure}
\label{sec:flow}
\subsection{General properties}
\label{subsec:GeneralProperties}

We here present the flow structure that results from the jets. We recall that we do not include the spiralling-in of the NS to the envelope as we take a circular orbit at $a=700 R_\odot$. As well, we include neither the NS gravity nor the self gravity of the ejected mass. We do include the spherically-symmetric gravity of the RSG as it is when we start the simulation (it does not change with time). We also do not include initial envelope rotation that we expect due to the spiralling-in of the NS into the envelope. Our flow maps emphasise the role of the jets in ejecting the envelope. 

Because we start with the NS already inside the envelope and the envelope is of an undisturbed RSG model, we must allow the binary system to have a relaxation time during which transient numerical features disappear. We find that this requires just over one orbital period of the NS. We therefore present the results only after two orbital periods $t=2 P_{\rm orb}= 3.55 \yr$.

\begin{figure*} 
\centering
\includegraphics[width=0.42\textwidth]{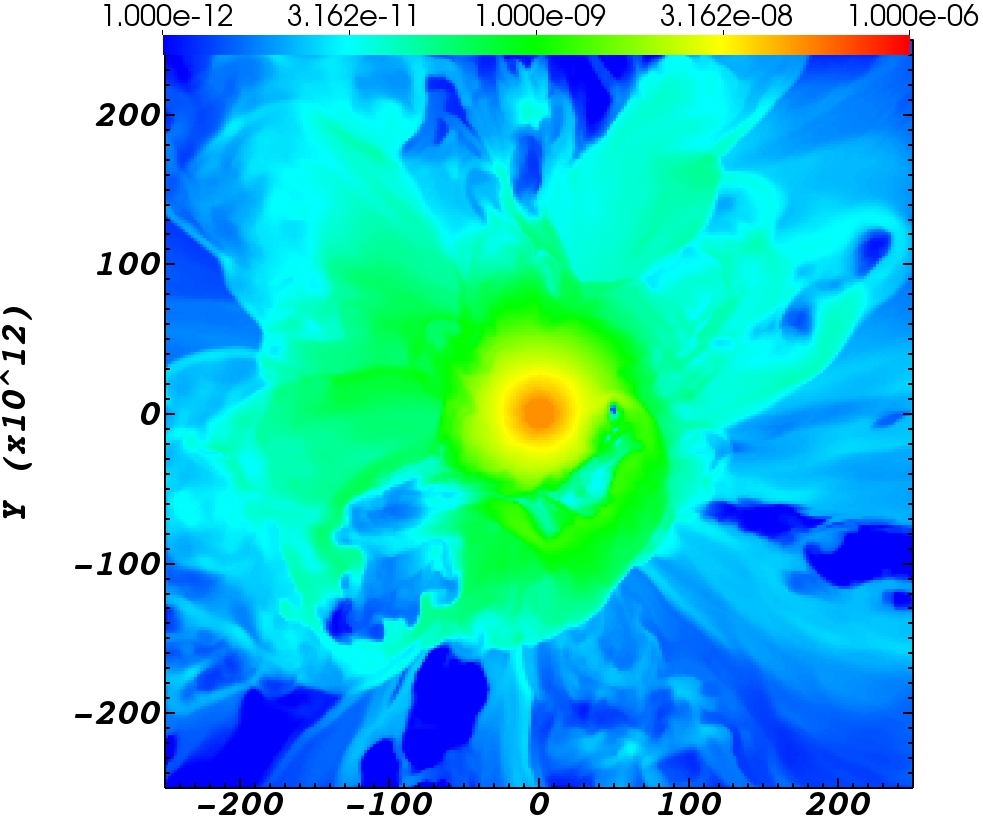} 
\includegraphics[width=0.42\textwidth]{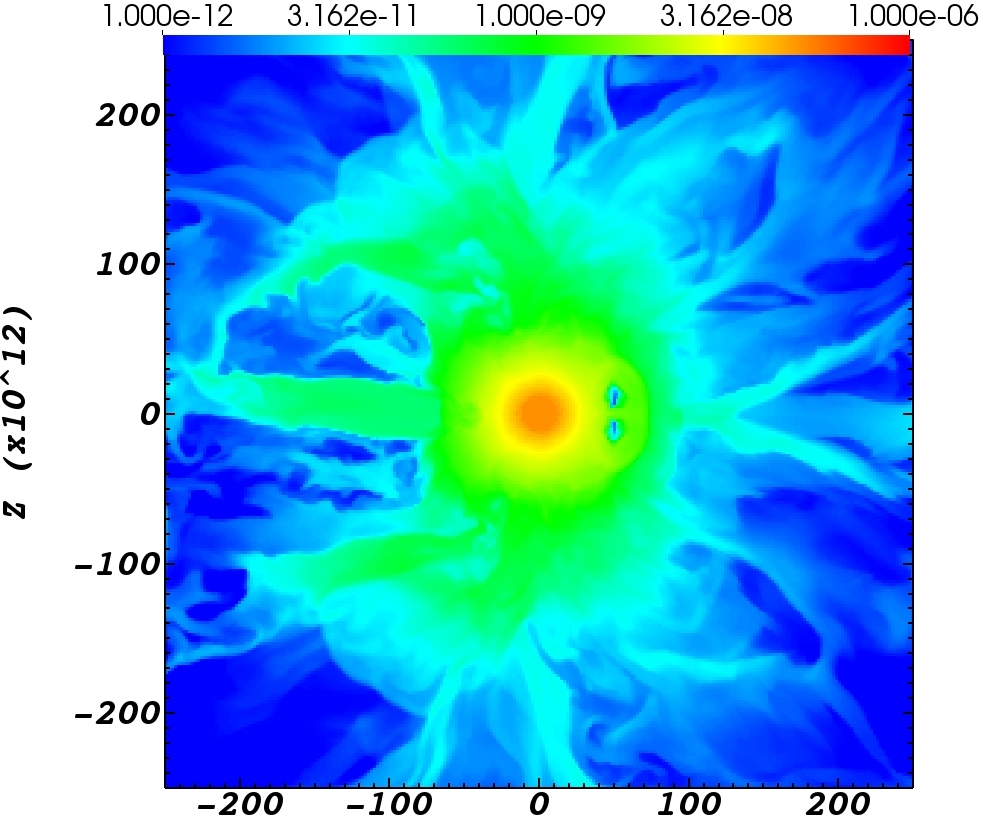}\\
\includegraphics[width=0.42\textwidth]{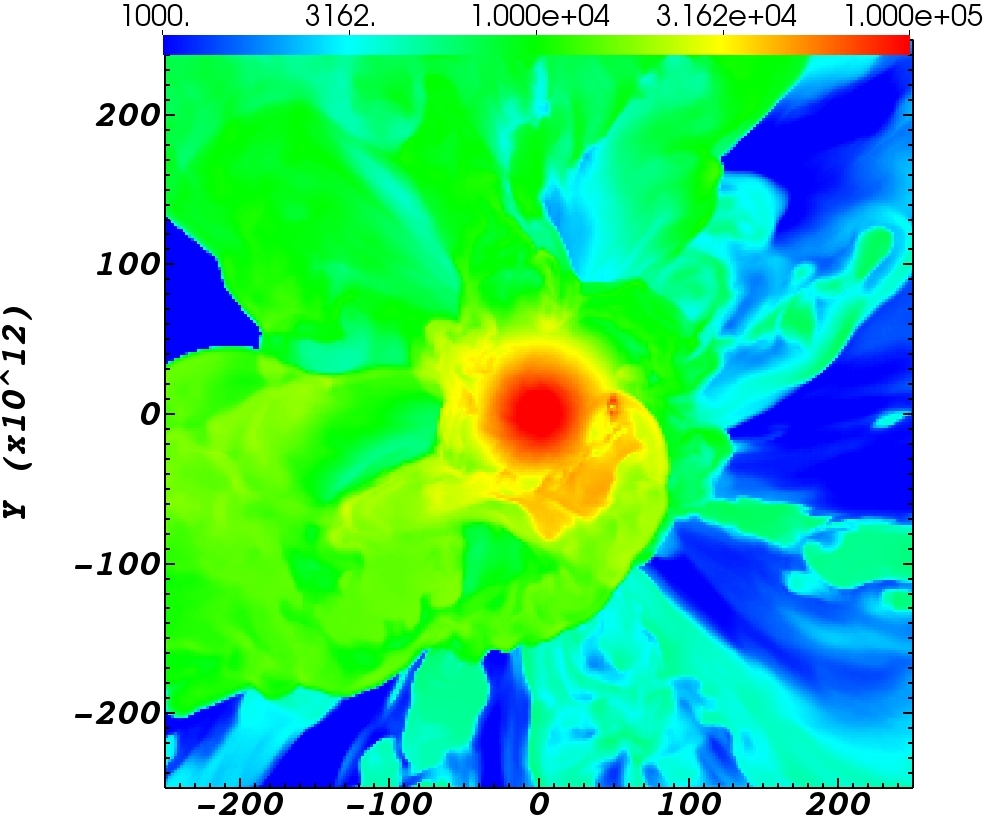} 
\includegraphics[width=0.42\textwidth]{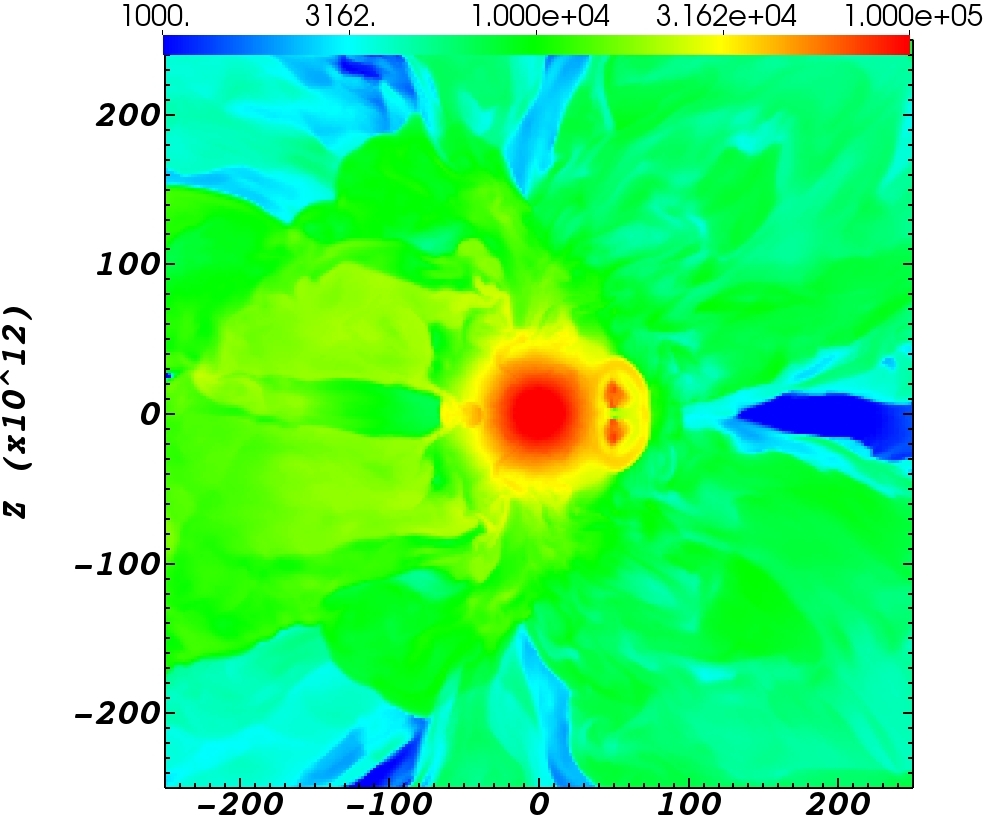}\\
\includegraphics[width=0.42\textwidth]{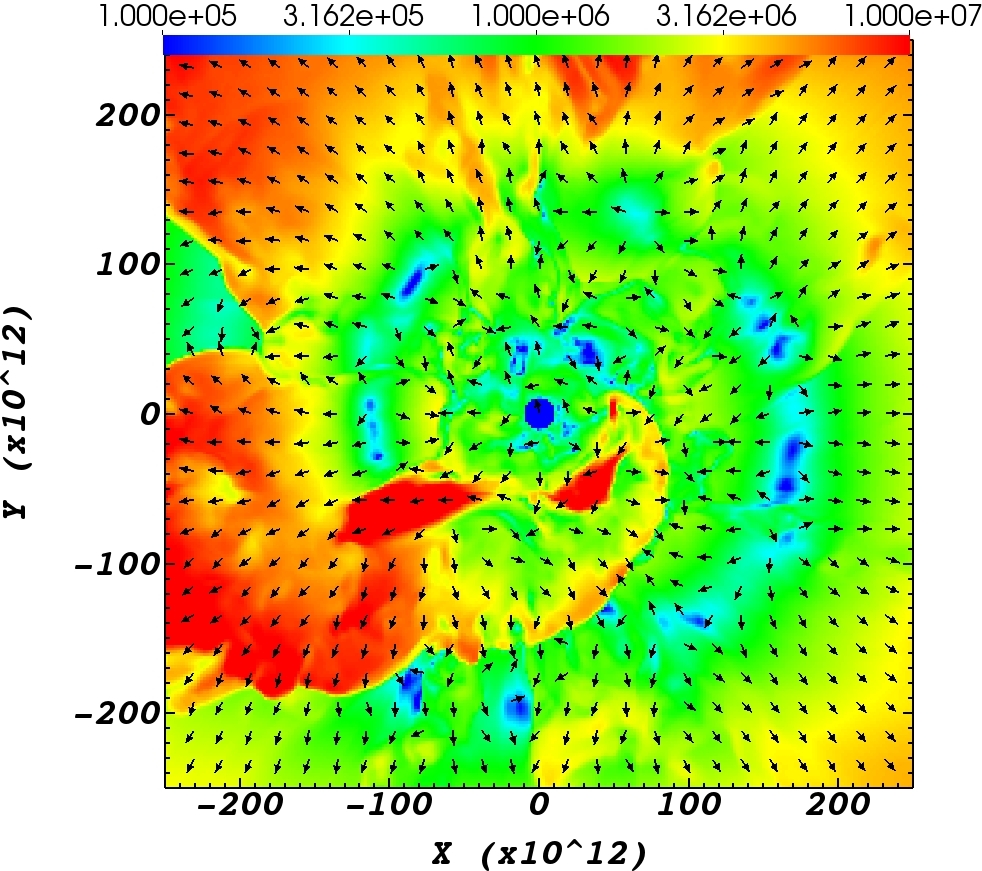}
\includegraphics[width=0.42\textwidth]{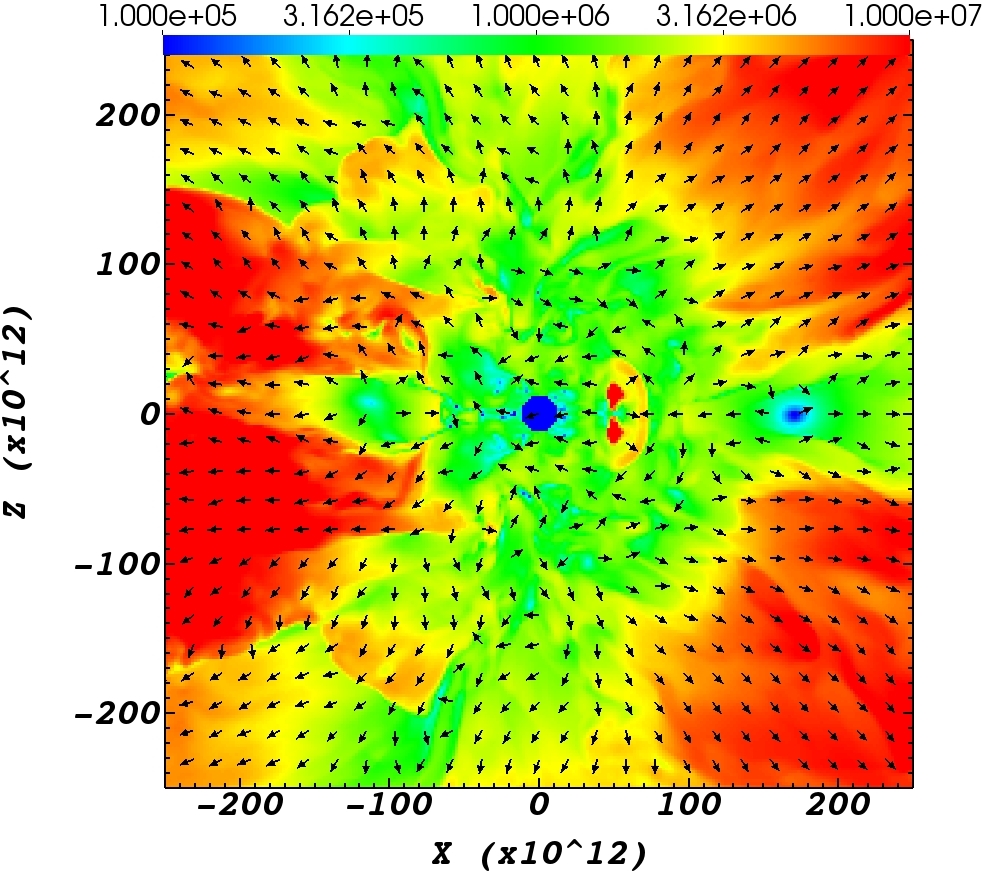}
\caption{Results of simulation Run40 at $t=2 P_{\rm orb}= 3.55 \yr$.  The axes of all panels are from $-250 \times 10^{12} \cm$ to $250 \times 10^{12} \cm$. Left column: maps in the plane $z=6 \times 10^{12} \cm = 0.86 L_{\rm L}$, where $L_{\rm L}$ is the length of the cone into which we inject each lobe, of the density (top panel; color bar from $10^{-12} \g \cm^{-3}$ to $10^{-6} \g \cm^{-3}$), temperature (middle panel; color bar from $1000 \K $ to $10^5 \K$) and velocity arrows on top of the velocity magnitude  (bottom panel; color bar from $10^{5} \cm \s^{-1}$ to $10^{7} \cm \s^{-1}$). The NS is rotating counter-clockwise.   
Right column: similar maps but in the $y=0$ meridional plane.
The two cones into which we inject the jets' energy appear on the right panels at $x=a=49\times 10^{12} \cm$ as two small regions on the two sides of the $z=0$ plane, of low density (blue regions), high temperatures (red regions), and high velocity (red regions).  }  
\label{fig:RUN40profiles}
\end{figure*}

In Fig. \ref{fig:RUN40profiles} we present the density, temperature, and velocity maps of simulations RUN40 at $t=2 P_{\rm orb}= 3.55 \yr$. At that time the NS is at $(x,y,z)=(49 \times 10^{12} \cm, 0, 0)$. The orbital plane is the $z=0$ plane and the jets' symmetry axis is perpendicular to the orbital plane and through the NS. In the left column we present these quantities in the plane $z=6 \times 10^{12} \cm = 0.86 L_{\rm L}$, where $L_{\rm L}$ is the length of the cone into which we inject each lobe (section \ref{subsubsec:interaction}), and in the right column we present these quantities in the meridional plane $y=0$ that contains the NS at that time. 

The not-too-high jets' energy of simulation RUN40 allows a large grid with a relatively high resolution (but still not sufficient to resolve the accretion process onto the NS). However, for the more realistic simulations RUN41, RUN41L, and RUN42L of higher jets' powers we have to use a smaller grid and larger cells (lower resolution). For that we present in Fig. \ref{fig:RUN40profiles} the results of simulation RUN40. 

\begin{figure*} 
\centering
\includegraphics[width=0.34\textwidth]{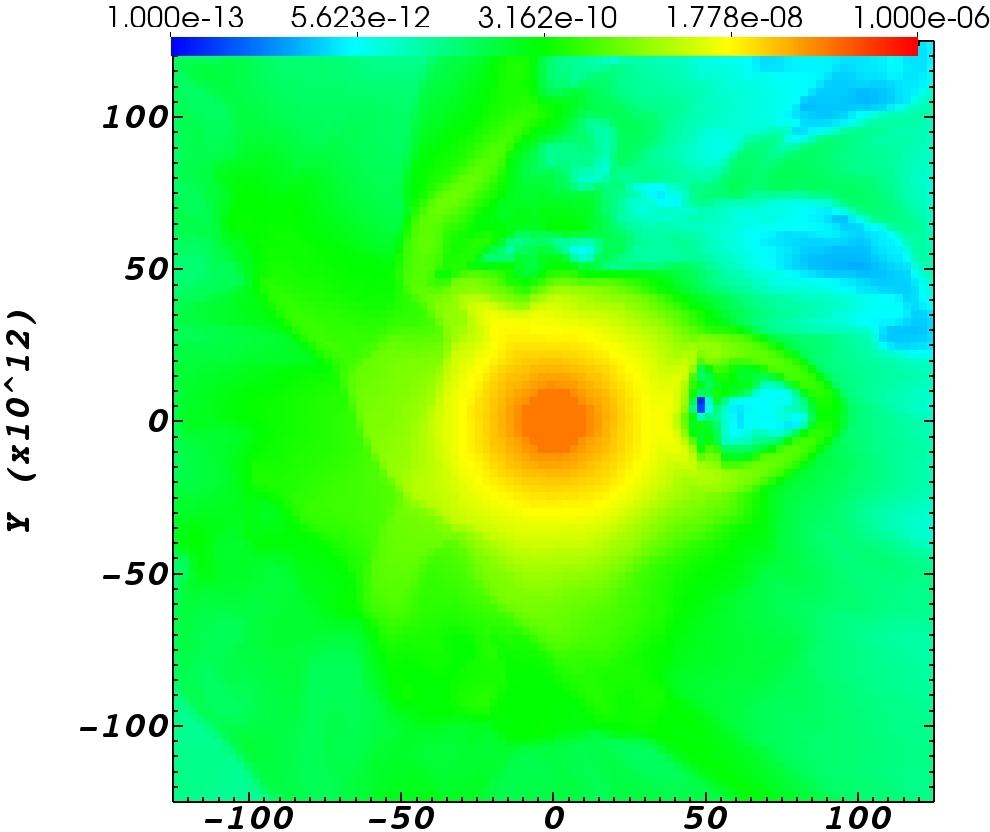}
\includegraphics[width=0.32\textwidth]{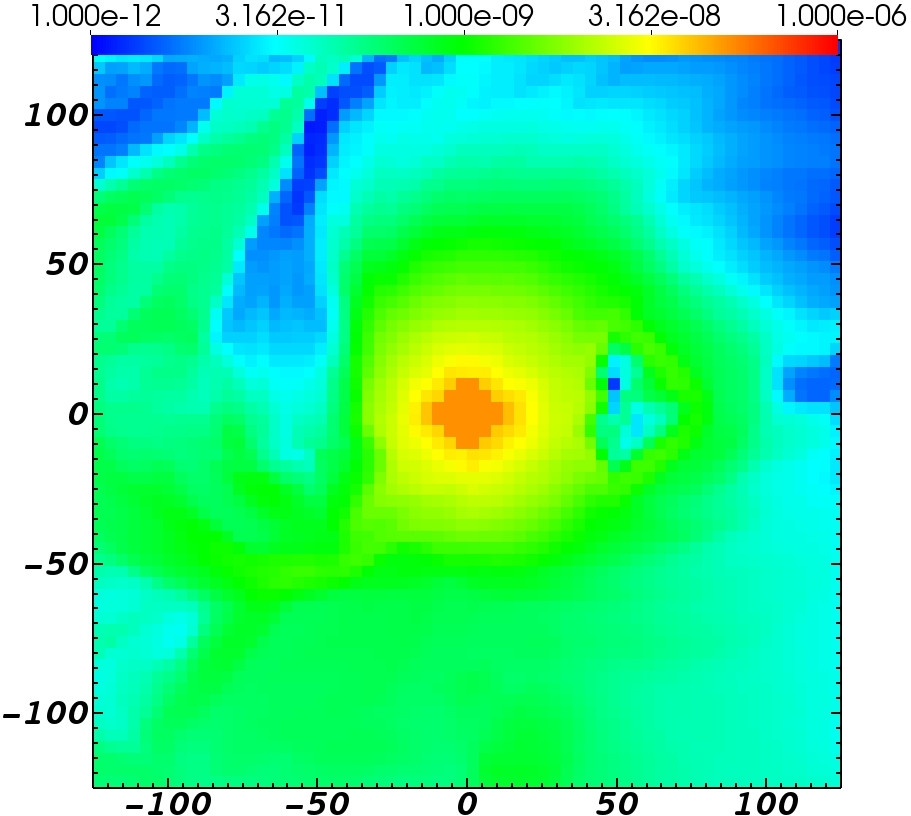} 
\includegraphics[width=0.32\textwidth]{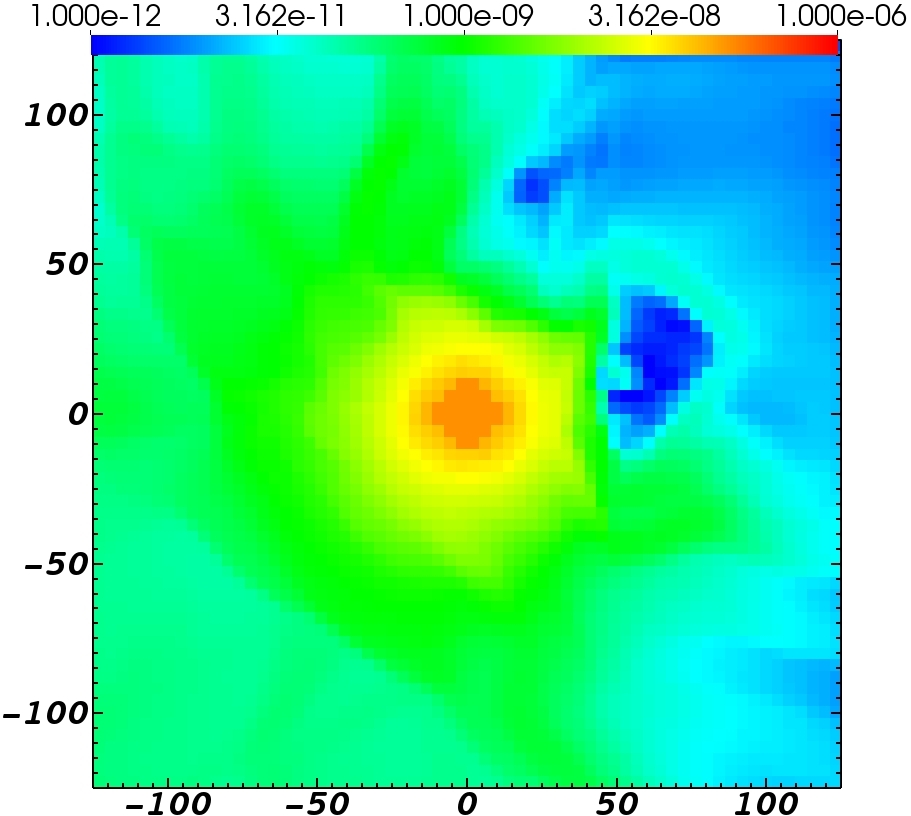}\\
\includegraphics[width=0.34\textwidth]{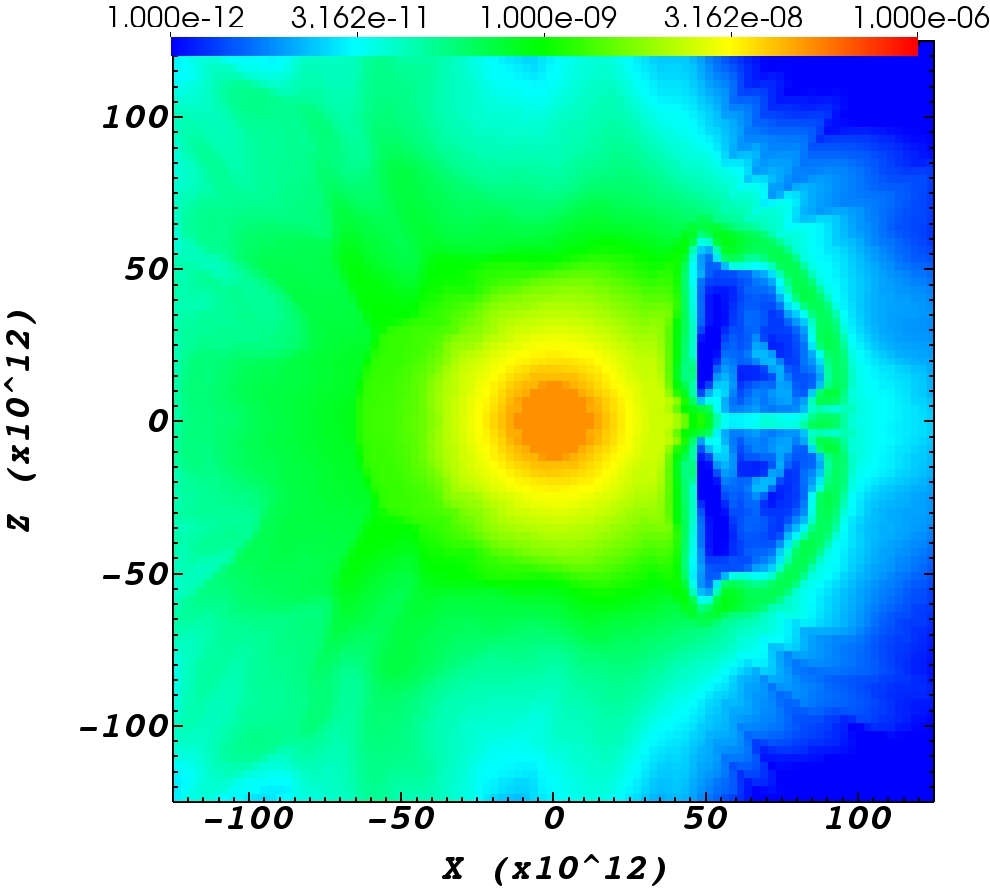}
\includegraphics[width=0.32\textwidth]{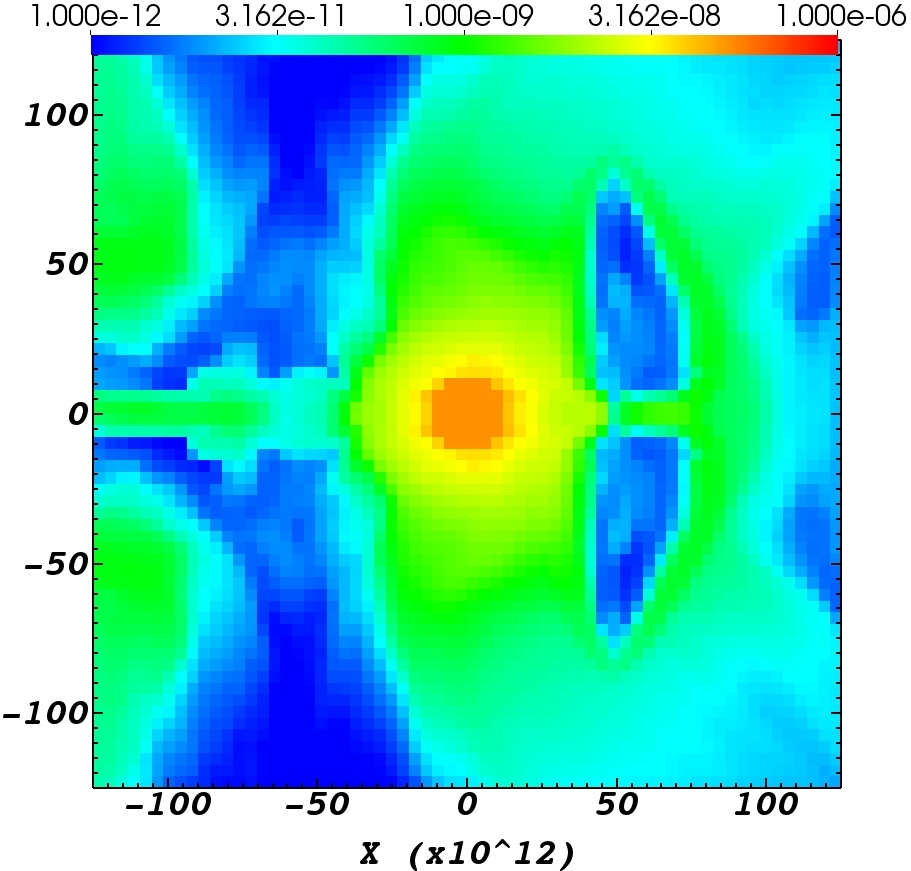}
\includegraphics[width=0.32\textwidth]{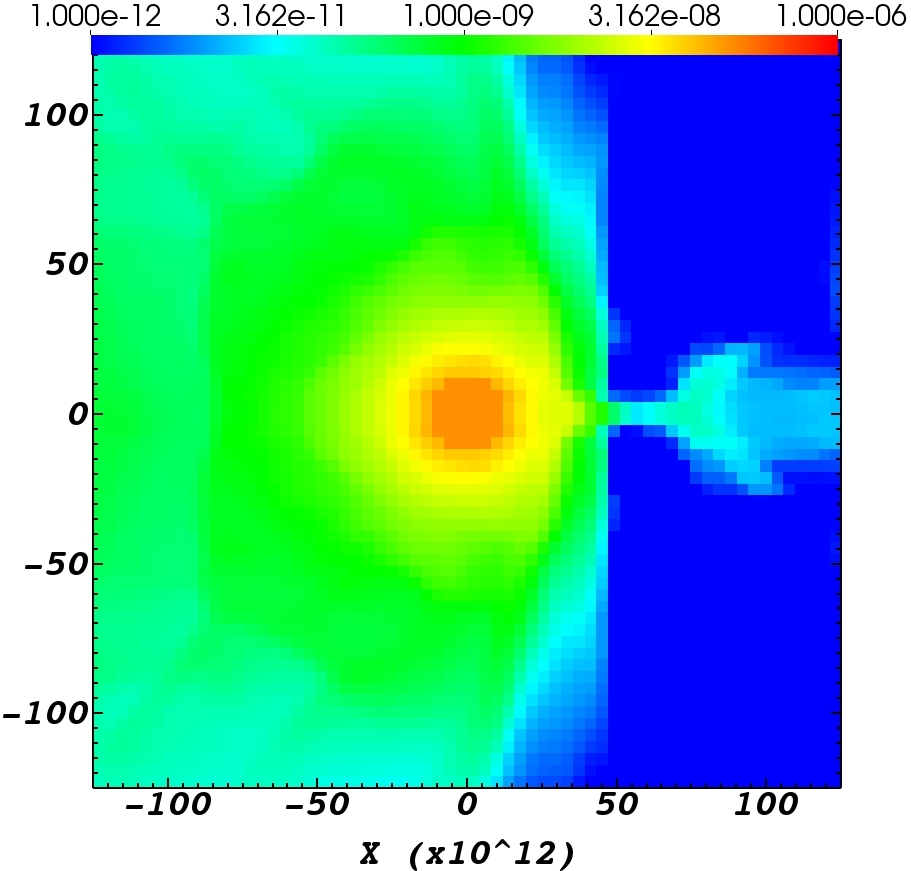}\\

\caption{The density maps of simulations RUN41 (left column), RUN41L (middle column), and RUN42L (right column), all at $t=2 P_{\rm orb}$. In the upper panels we present the density maps in the plane $z= 0.86 L_{\rm L}$, and in the lower panels in the meridional plane $y=0$ that contains the NS that is at $(x,y,z)=(49 \times 10^{12} \cm, 0, 0)$. 
Colour bars are from $10^{-12} \g \cm^{-3}$ to $10^{-6} \g \cm^{-3}$.
The central square extending to $x=\pm 12 \times 10^{12}$ and $z=\pm 12 \times 10^{12}$ is the inert core of the simulation. It appears square and not circle because of the low numerical resolution. Note that much larger low density bubbles (cocoons) that the jets inflate relative to the case in Fig. \ref{fig:RUN40profiles}.  The axes of all panels are from $-125 \times 10^{12} \cm$ to $125 \times 10^{12} \cm$. }  
\label{fig:RUN41vs42}
\end{figure*}

\begin{figure*} 
\centering
\includegraphics[width=0.34\textwidth]{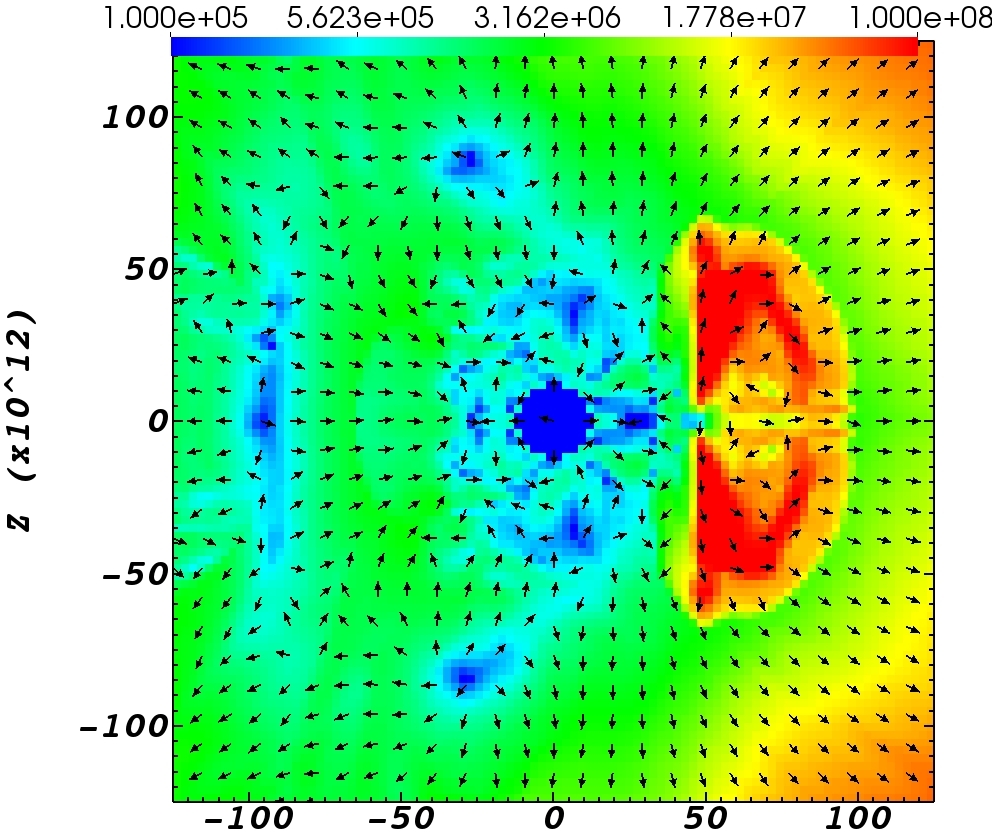}
\includegraphics[width=0.32\textwidth]{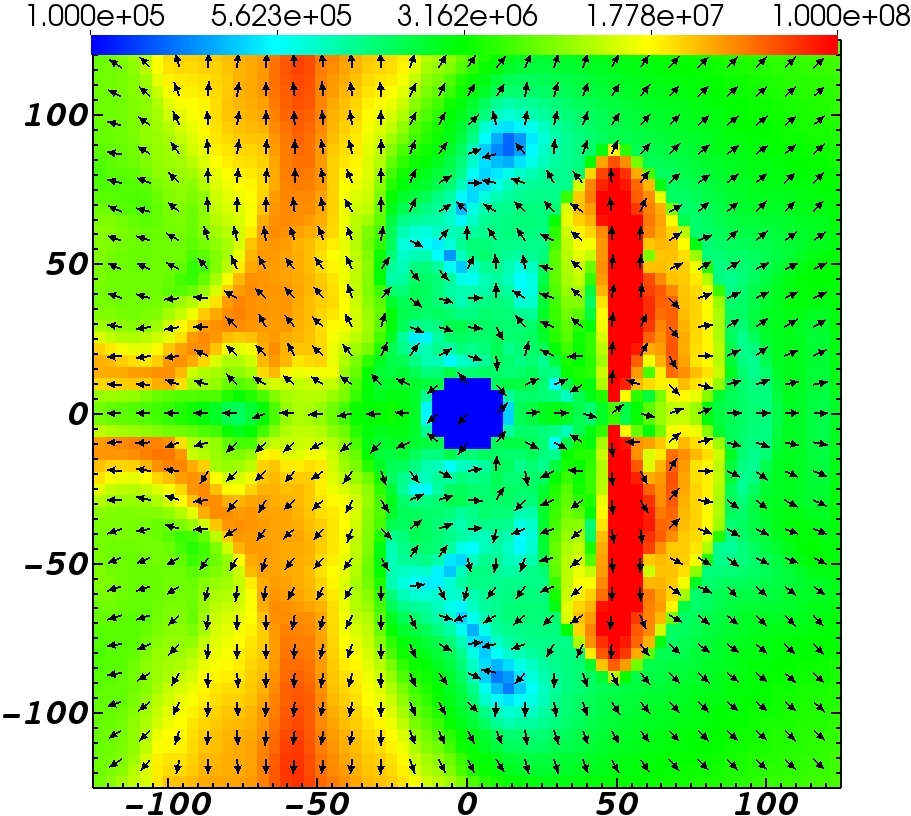}
\includegraphics[width=0.32\textwidth]{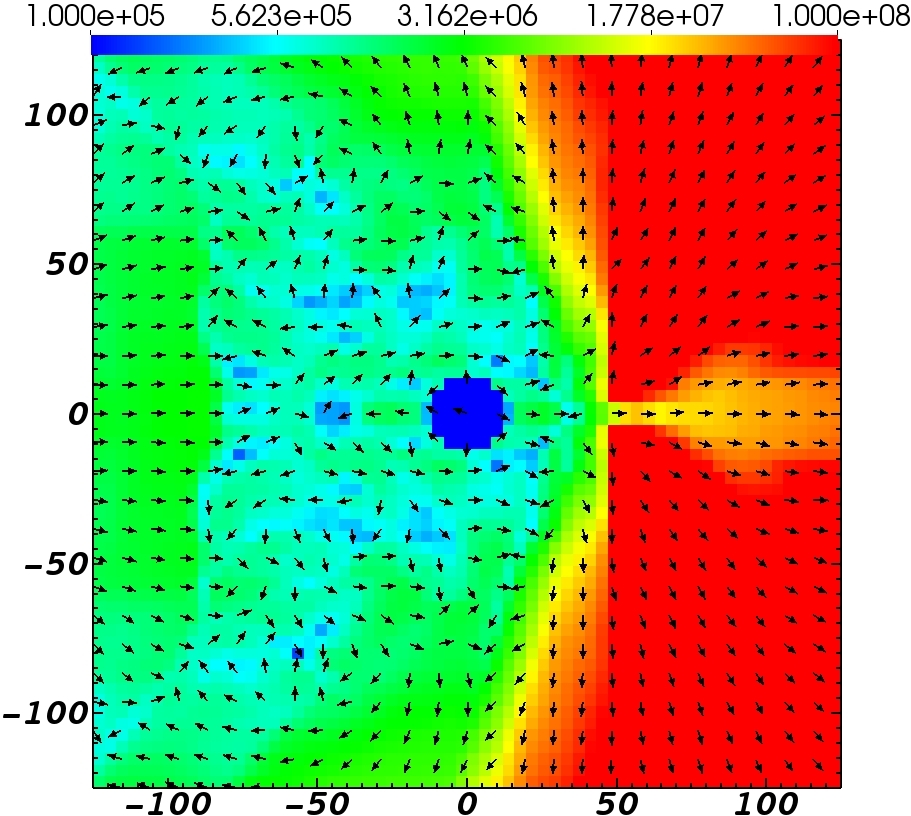}\\
\includegraphics[width=0.34\textwidth]{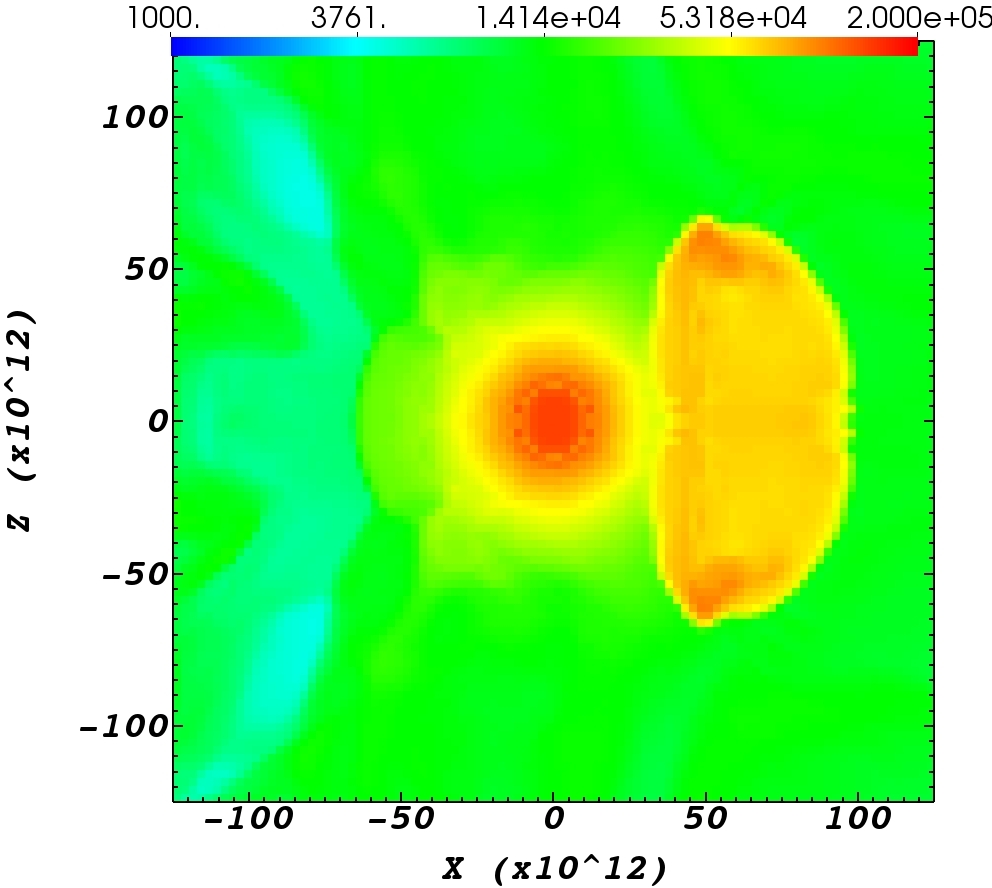}
\includegraphics[width=0.32\textwidth]{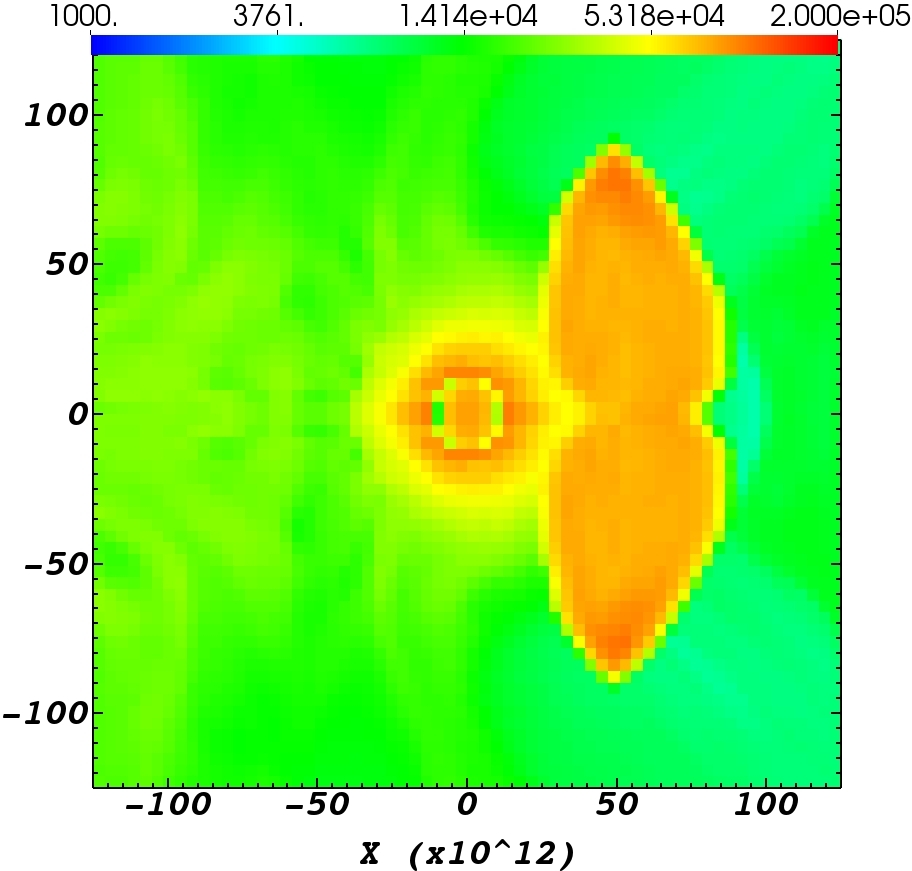}
\includegraphics[width=0.32\textwidth]{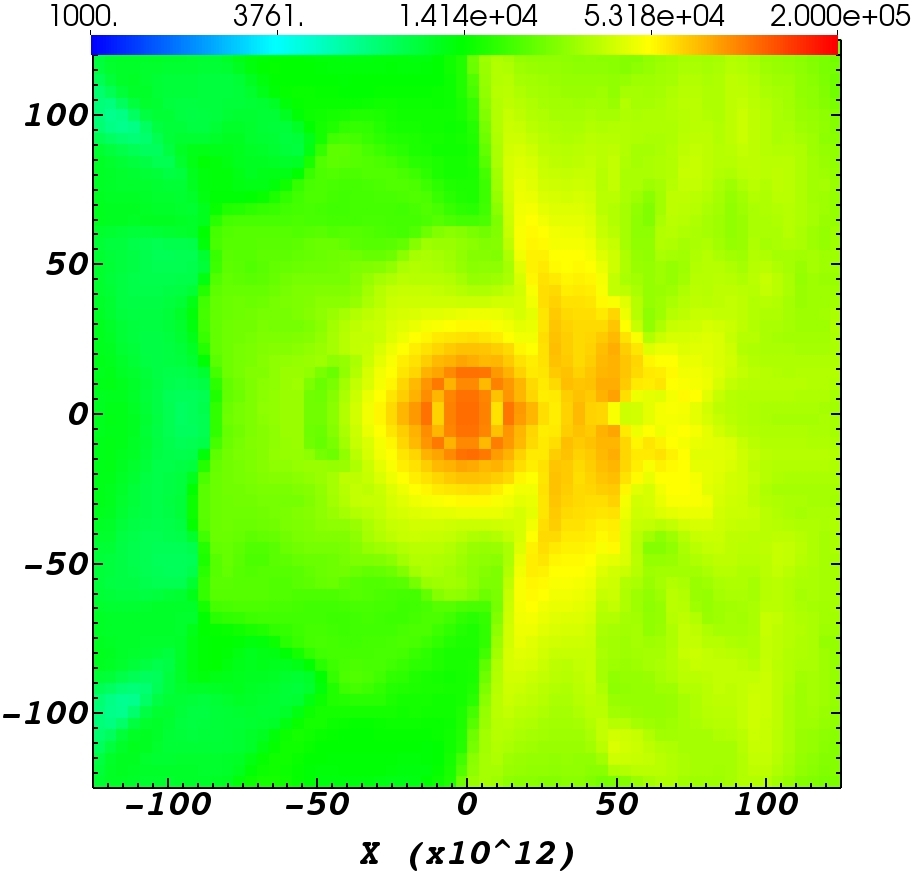}\\
\caption{The velocity (upper panels; color bar from $10^{5} \cm \s^{-1}$ to $10^{8} \cm \s^{-1}$) and temperature (lower panels; color bar from $1000 \K $ to $2 \times 10^5 \K$) maps in the meridional plane $y=0$ of simulations RUN41 (left column), RUN41L (middle column), and RUN42L (right column), all at $t=2 P_{\rm orb}$.
Note that the velocity scale is an order of magnitude above that of Fig. \ref{fig:RUN40profiles}.  The axes of all panels are from $-125 \times 10^{12} \cm$ to $125 \times 10^{12} \cm$.
}   
\label{fig:RUN41vs42Vel}
\end{figure*}

In Figs. \ref{fig:RUN41vs42} and \ref{fig:RUN41vs42Vel} we present the flow of simulations RUN41 (left column) RUN41L (middle column) and RUN42L (right column). In Fig. \ref{fig:RUN41vs42} we present the density map in the plane $z= 0.86 L_{\rm L}$ (top panels), and in the $y=0$ meridional plane (lower panels), while in Fig. \ref{fig:RUN41vs42Vel} we present the velocity (upper panels) and temperature (lower panels) maps in the meridional plane $y=0$. We present these panels at $t=2 P_{\rm orb}$.  

 We note the differences between RUN41 (left columns) and RUN41L (middle columns of Figs. \ref{fig:RUN41vs42} and \ref{fig:RUN41vs42Vel}). In RUN41L the low-density high-temperature bubbles that the jets inflate are narrower and more elongated along the $z$-direction from the NS than the bubbles in RUN41; the NS is at $(x,z)=(49\times 10^{12} \cm, 0)$ in the lower panels of Fig. \ref{fig:RUN41vs42} and in all panels of Fig. \ref{fig:RUN41vs42Vel}. The bubbles are the blue bipolar lobes in the two left panels in the lower row of Fig. \ref{fig:RUN41vs42} and the red bipolar lobes in the two left columns Fig. \ref{fig:RUN41vs42Vel}.  We also note that in RUN41L the velocities are higher and densities are lower on the left side of the RSG core, Namely, the region where the NS was half a period earlier. We attribute these differences to the limiting resolution that we have in the jet injection region. We inject each jet into a cone of length $7\times 10^{12} \cm$. In RUN41 the size of the cells in that region is $2.6 \times 10^{12} \cm$ while in RUN41L it is $3.9 \times 10^{12} \cm$. Namely, we resolve the jet-injection cone with less than 3 and less than 2 cells in RUN41 and RUN41L, respectively. These are low resolutions, but we are on the edge of our computer capabilities. We return to discuss our resolution in section \ref{sec:Uncertainties}.

We expect the jets in reality to be stronger than what our most energetic simulation RUN42L is (section \ref{sec:Feedback}). Nonetheless, we can learn the following from our simulations.

\subsection{Grazing envelope evolution (GEE)}
\label{subsec:Grazing}

We find that for the low-energy simulations RUN39, RUN40 and RUN41L the jets do not break out from the envelope (i.e., choked jets). We see this as the blue lower density opposite zones where we inject the jets' energy that are surrounded by the dense envelope gas (in green; upper right panel of Fig. \ref{fig:RUN40profiles} and lower left panel of Fig. \ref{fig:RUN41vs42}). In the more realistic simulation RUN42L the jets break out from the envelope and remove a very large portion of the outer envelope.
The initial mass between the inert core at $r=176R_\odot$ and the radius of $r=700 R_\odot$ is $5.8 M_\odot$. At the end of simulation RUN39 (about 4 years) the mass in that zone is $4.9 M_\odot$, while in simulation RUN42L the mass in this zone after 4 years is $2.7 M_\odot$. At the end of the simulations RUN42L (about 6.3 year) the mass in this zone is only $2.4 M_\odot$.  
We recall that we include gravity in the simulations as it is at $t=0$. Therefore, considering the mass of the inert core of $5.64 M_\odot$, the assumption of constant gravity is sufficiently accurate for simulation RUN39. However, in simulation RUN42L, because of the substantial removal of mass, the gravity in reality would be weaker, and as a consequence of that jet break-out would be easier.  

In simulation RUN42L the jets remove most of the envelope gas outside the orbit of the NS. If this holds for more realistic simulations that include more energetic jets and the spiralling-in process as well, it might have an implication to the onset of the CEE in hinting that at the onset of the CEE the jets remove the entire envelope outside the NS orbit. Namely, it is possible that in the outer regions of the envelope the system would experience the GEE (for hydrodynamical simulations of the GEE but with a main sequence companion rather than with a NS companion see,e.g., \citealt{Shiber2018, Shiberetal2019, LopezCamaraetal2021}). 

 Note that to remove the envelope outer to the NS orbit, the minimum energy requirement on the jets decreases with increasing radius. This is because both the outer envelope mass and the gravity decrease with increasing radius. The condition on the jets' power we find here to remove the envelope, i.e., $\dot E_{\rm 2j} \ga 10^{42} \erg \s^{-1}$, is true at the orbital radius of $a=700 R_\odot$ that we assume here. At earlier times, when the NS spirals-in at larger radii inside the envelope, lower-power jets can remove the envelope. The value of $\dot E_{\rm 2j}$ increases as the NS spirals in. As well, to perform the GEE the spiralling-in should be slow, which in turn implies that the RSG envelope is rapidly rotating (to reduce tidal forces). We require therefore the NS to substantially spin-up the envelope before it enters the envelope for a GEE to be significant.

\subsection{Convection energy transport}
\label{subsec:Convection}

From the velocity maps (lower panels of Fig. \ref{fig:RUN40profiles} and upper panels of Fig. \ref{fig:RUN41vs42Vel}) we learn that the jets induce flow patterns that include inflow and outflow with respect to the center of the RSG. This flow patter increases the flow in the envelope due to the strong convection. The convection in cool giants can be very efficient in transporting energy out, whether recombination energy or the energy that a companion deposits in the envelope during a CEE (e.g., \citealt{Sabachetal2017, Gricheneretal2018, WilsonNordhaus2019, WilsonNordhaus2020, WilsonNordhaus2022}). Our results hint at the possibility that the flow that the jets induce in the envelope makes the convective energy transport more efficient even. 

 The exact flow pattern, including the turbulence, that the jets induce in the envelope depends on their power. Even the weak jets of simulation RUN40 manage to induce turbulence in the entire envelope, as we see in the lower two panels of Fig. \ref{fig:RUN40profiles}. The typical amplitudes of the turbulence is $v_{\rm tur} \approx {\rm few} \times 10 \km \s^{-1}$ (yellow and orange colors in the two lower panels of Fig. \ref{fig:RUN40profiles}), although most of the gas flows at $\simeq 10 \km \s^{-1}$ (green in the figure). We see the turbulence on both sides of the velocity plots in Fig. \ref{fig:RUN40profiles}, namely, not only near the NS.

 The medium-power jets of RUN41 (upper-left panel of Fig. \ref{fig:RUN41vs42Vel}) and RUN41L (upper-middle panel) induce turbulence that are faster by about a factor of 3, as the green color in Fig.  \ref{fig:RUN41vs42Vel} corresponds to a velocity of $\simeq 30 \km \s^{-1}$. This can be understood from the ten times higher power of the jets in RUN41 with respect to RUN40. The ten times more powerful jets of RUN41 and RUN41L induce a more pronounced outflow, i.e., a faster ordered outflow than that in RUN40. 

 The flow pattern of the high-power jets of RUN42L has qualitative differences. The reason is that the jet-inflated bubbles of RUN42L break out from the envelope, allowing the high-energy gas to escape. We can see this escape of energy in the lower panels of Fig. \ref{fig:RUN41vs42Vel}. The temperatures in the high-temperature zones (red) of RUN42L are lower than those of RUN41 and RUN41L. The reason is that the hot bubbles that the jets in simulation RUN42L inflate escape and cool adiabatically. We see this flow pattern in the velocity map in the upper-right panel of Fig. \ref{fig:RUN41vs42Vel}. The more energetic jets of RUN42L accelerate a very large region to high velocities, i.e., the large red region on the right. This gas escapes from the envelope. Therefore, less energy is deposited into the bound envelope. If we look at the left regions of the velocity maps, where the NS launched the jets half an orbit earlier, we see that the velocities of RUN42L are lower than those of RUN41L in that region. The reason is that when the NS was in that region, the bubbles that the jets of RUN42L inflated escaped from the envelope. It seems that when the jets manage to break out, the turbulence in the still-bound envelope becomes weaker, even though the jets are stronger, because the high-energy gas escapes from the envelope.

The main conclusions from this section (the three subsections) carry on to future numerical simulations of CEE with a NS/BH companion are that the simulations should include energy transport by convection (including the jet-induced downward-flow and upward-flow) and that researchers should be alert of the possibility of GEE at the beginning of the spiralling-in process in the envelope. 

\section{The negative jet feedback coefficient}
\label{sec:Feedback}

\begin{figure*} 
\centering
\includegraphics[width=0.96\textwidth]{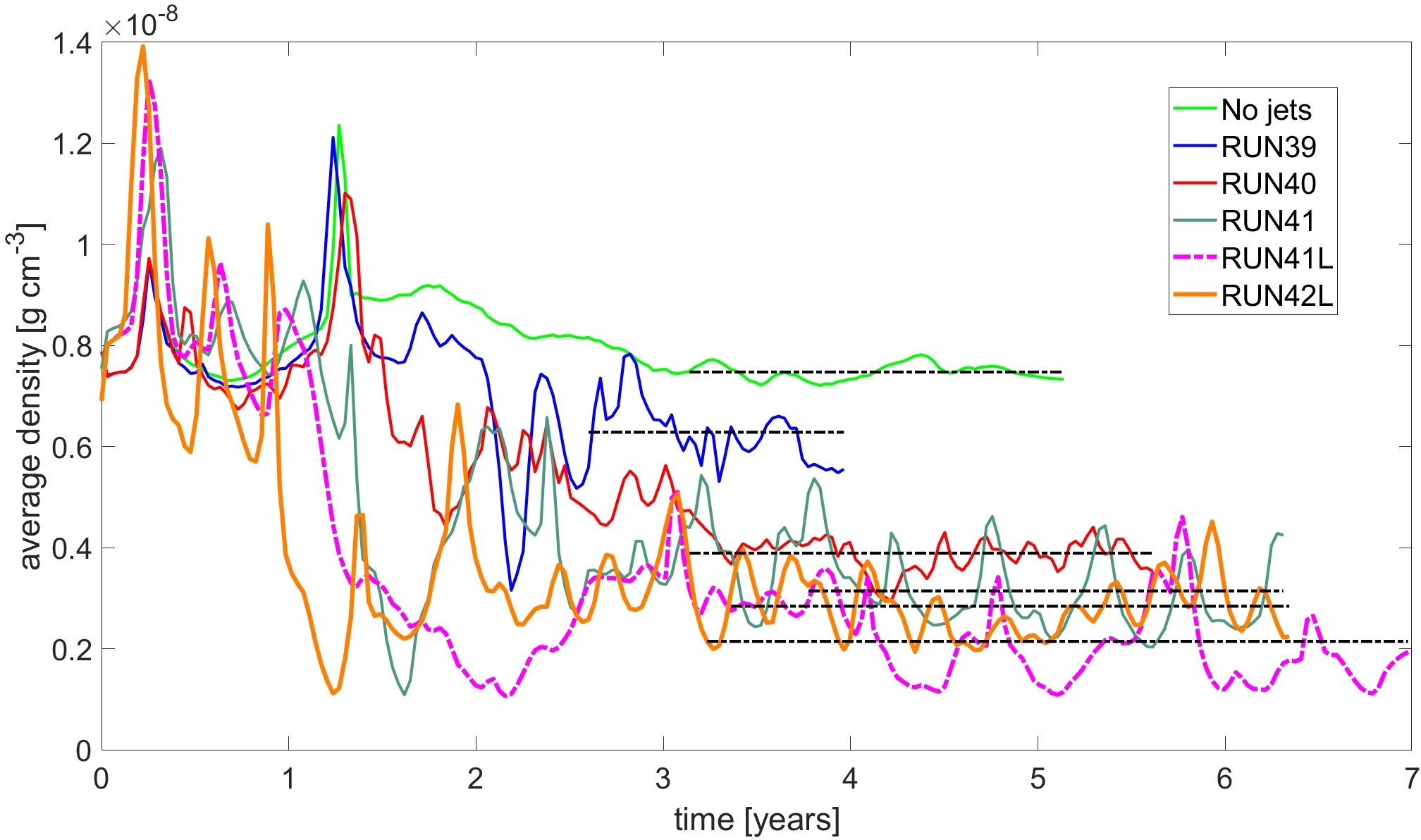}
\caption{The average density $\bar \rho (t)$ over a thin spherical shell of radius $r=a=700 R_{\odot}$ as a function of time. The dashed-dotted horizontal line of each of the five cases with jets and one case without jets marks the time-averaged density $<\bar \rho (t)>$ over a time periods as the dashed-dotted line marks. 
}  
\label{fig:AverageD}
\end{figure*}

We follow \cite{Gricheneretal2021} in finding the negative jet feedback coefficient $\chi_j$,which is the factor by which the operation of the jets reduces the accretion rate. We take this value to be the factor by which the jets reduce the density along the NS orbit $q_\rho$, namely, 
\begin{equation}
\chi_{\rm j} = q_\rho \equiv \frac {< \bar \rho (t) > }{\rho_{\rm 0}} ,
\label{eq:ChiJ}
\end{equation}
where $\bar \rho (t)$ is computed as the average density over a thin spherical shell centred at the origin of the grid (centre of the RSG) and with a radius equals to the orbital radius of the NS, $< \bar \rho (t) >$ is its time-averaged value, and $\rho_0$ is the density in the same region in the unperturbed envelope. For the value of $\rho_0$ we take the average value of the density in the simulation without jets (green line in Fig. \ref{fig:AverageD}) after the model has relaxed (upper dashed-dotted horizontal line in Fig. \ref{fig:AverageD}). This density is about equal to the initial density in the unperturbed 1D stellar model we use (see below). 
We do not take the density within an accretion radius of the NS because we neglect the self gravity of the envelope and the NS's gravity, and because we do not follow the exact jet-envelope interaction. 
In Fig. \ref{fig:AverageD} we present the variation with time of $ \bar \rho (t)$ for the different cases. 

We calculate the time average $<\bar \rho (t) >$ over the time that we present by the horizontal dashed-dotted black lines in Fig. \ref{fig:AverageD}. We start the time averaging after the relaxation of the initial large perturbations. The density continues to fluctuate during the entire simulations, and so we average over few fluctuations.
We justify this approach by the behavior of the simulation without jets. In that simulation there are large density fluctuations in the first three years (greed-dashed-dotted line in Fig. \ref{fig:AverageD}). However, after that time period the fluctuations decrease and the average value is very close to the initial density of the 1D model near the NS orbit, i.e., $\rho_{\rm 1D} \simeq < \bar \rho_{\rm no-jets}(t) > \equiv \rho_0 =0.74 \times 10^{-8} \g \cm^{-3}$ (see also Fig. \ref{fig:testcase}).  We list the values of $q_\rho$ in the sixth column of Table \ref{Table:cases}.  

With the density reduction due to the jets the actual accretion rate onto the NS is then 
\begin{equation}
\dot M_{\rm acc,j} = \chi_{\rm j} \dot M_{\rm acc,0}=  \chi_{\rm j} \xi \dot M_{\rm BHL,0},  
\label{eq:AccretionRate}
\end{equation}
where $\xi$ is the ratio of the accretion rate to that of the analytical BHL value $\dot M_{\rm BHL}$ (\citealt{HoyleLyttleton1939}; \citealt{BondiHoyle1944}).
According to hydrodynamical simulations in most cases $\xi \simeq 0.1-0.5$(e.g., \citealt{Livioetal1986} and \citealt{RickerTaam2008} who did not include jets, and \citealt{Chamandyetal2018} who included jets), although some simulations deduce smaller values of $\xi < 0.1$ (e.g., \citealt{MacLeodRamirezRuiz2015a}; \citealt{MacLeodRamirezRuiz2015b}). 

The jets themselves carry a faction of $\eta \simeq 0.1$ (down to $0.05$) of the accretion power. Neutrino carry most of the accretion power as it is the neutrino cooling process that allows the high accretion rate.  Overall, the power of the jets is according to equation (\ref{eq:E2jzeta}) with  
\begin{equation}
\zeta \equiv \eta \chi_{\rm j} \xi \simeq 0.005 \chi_{\rm j} - 0.05 \chi_{\rm j}, 
\label{eq:Zeta}
\end{equation}
where in the second equality we substituted $\xi \eta \simeq 0.005-0.05$.

\begin{figure} 
\centering
\includegraphics[width=0.42\textwidth]{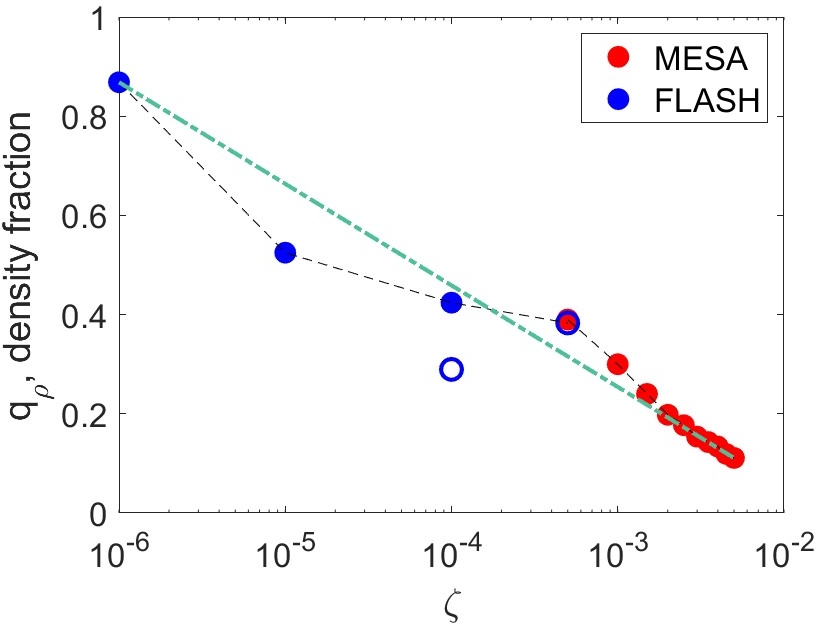}
\caption{The magnitude of the jet feedback mechanism from the 1D simulations of \cite{Gricheneretal2021} (red dots) and our new 3D simulations in blue dots. The two open blue-dots are of simulations RUN41L and RUN42L that are less reliable because of their lower resolution. Note that the point of simulation RUN42L almost coincides with the upper point of the 1D simulations.   
The green dashed-doted line is a fit through all the results
according to equation \ref{eq:E2jzeta}. 
} 
\label{fig:jetspower}
\end{figure}

As for each simulation we predetermine the value of $\zeta$ (Table \ref{Table:cases}) and derive the density ratio $q_\rho$ (Table \ref{Table:cases}), by taking $\chi_{\rm j} = q_\rho$ we can determine which value of $\chi_j$ fulfils  equation (\ref{eq:Zeta}). 
These simulations we consider to be self consistent according to the jet feedback mechanism and the values of $\eta \xi$ we adopt here. 
 We cannot simulate high values of $\zeta$ because of numerical limitations. We therefore present in Fig. \ref{fig:jetspower} results for both our new 3D simulations (blue upper dots) and results from \cite{Gricheneretal2021} for their 1D simulations with a spiralling-in to orbital radius of $100 \AU$ (lower red dots). Note that our right-most point (RUN42L) almost coincides with the upper point of \cite{Gricheneretal2021}, as seen by the red dot inside the blue circle.   

We comment on the uncertainty in our results here, and then in section \ref{sec:Uncertainties}. The envelope density for the high-power simulations fluctuate around the mean (the dashed-dotted horizontal lines in Fig. \ref{fig:AverageD}) by an amplitude of $\Delta \rho \simeq 0.1 \times 10^{-8} \g \cm^{-3}$. The large fluctuations introduce uncertainties to the calculated average density. This adds to the uncertainties that the limited resolution that we can allow introduces. We see this in the different values that simulations RUN41 and RUN41L give. Nonetheless, despite these uncertainties in the individual simulations, we have confident in our results because of the common trend of our 5 simulations and the 1D results as we present in Fig. \ref{fig:jetspower}. 

We manage to crudely fit the two sets of simulations with a line 
\begin{equation}
\chi_{\rm j} = -0.2 \log \zeta -0.34.  
\label{eq:ChiFit}
\end{equation}
Substituting equation (\ref{eq:Zeta}) we obtain the equation for the negative jet feedback coefficient
\begin{equation}
5 \chi_{\rm j} + \log \chi_{\rm j} = -\log (\eta \xi)-1.7  .  
\label{eq:ChiFinal}
\end{equation}
For $\eta \xi = 0.01$, $\eta \xi = 0.02$ and $\eta \xi = 0.05$, as examples, we obtain , $\chi_{\rm j}=0.2$, $\chi_{\rm j}=0.16$, and $\chi_{\rm j}=0.11$ respectively.
Overall, we find 
\begin{equation}
\chi_{\rm j} \simeq 0.1-0.2 \quad {\rm for} \quad \eta \xi = 0.01-0.05.  
\label{eq:ChiFinalValue}
\end{equation}
Despite the uncertainties in the 1D simulations of \cite{Gricheneretal2021} and the uncertainties and density fluctuations in our new 3D simulations, we consider our estimate in equation (\ref{eq:ChiFinalValue}) to be reasonable. 
  
\section{Uncertainties}
\label{sec:Uncertainties}
 
 The challenging numerical simulations that require large computational resources and the oscillatory nature of the average density that we see in  Fig. \ref{fig:AverageD}, which we believe is a physical effect, introduce uncertainties to the results. We now discuss these.

 In section \ref{subsec:GeneralProperties} we discussed the differences in the flow patterns between RUN41 and RUN41L that we present in the left columns and middle columns, respectively, of Figs. \ref{fig:RUN41vs42} and \ref{fig:RUN41vs42Vel}. 
We attributed the differences mainly to the lower resolution of RUN41L, where we resolve each jet-injection cone with less than 2 cells. However, there is more to that. The general behavior of the flow is semi-oscillatory, as we clearly see by the average density plots that we present in Fig. \ref{fig:AverageD}.
The amplitude of the semi-oscillations of the average density depend on the resolution and jet power, which we vary in this study, and on the way we inject jets, which we do not vary in the present study. From Fig. \ref{fig:AverageD} we learn that RUN41, RUN41L, and RUN42, have large amplitudes. During the time $t> 2 P_{\rm orb} = 3.55 \yr$ the maximum average density is about twice the lowest density. The largest variations from minimum to maximum average density are in RUN41L (dashed-pink line). These large oscillations of RUN41L might lead to a higher mass loss rate, accounting for the lowest point of RUN41L in Fig. \ref{fig:jetspower}. 

 We conclude that there are two types of uncertainties in the values of $q_{\rho}$ that we plot in Fig. \ref{fig:jetspower} and in the value of the negative jet feedback coefficient that we give in equation (\ref{eq:ChiFinalValue}). The first type of uncertainties is due to the numerical setting, in particular due to the way by which we inject the jets to the grid, the numerical resolution, and due to the orbit of the NS. 
In the present study of 3D hydrodynamical simulations we inject the jets in two cones with varying resolutions between different simulations, and assume a constant circular orbit. In the 1D \textsc{mesa} stellar evolution simulations by \cite{Gricheneretal2021} the energy was deposited in spherical zones for an in-spiralling NS. Despite these differences in the input, in the grid symmetry, and in the type of simulations (hydrodynamics versus stellar evolution), there is a nice trend between the 1D and 3D simulations as the green dashed-doted line in Fig. \ref{fig:jetspower} shows.   

 The second type of uncertainties results from the large fluctuations that we discussed above. We think that these are real and that they result from real flow instabilities. However, the exact time variations of these large fluctuations depend also on the numerical setting that introduce the perturbations that instabilities in the flow amplify. We suggest by our results that the large fluctuations introduce larger uncertainties in the final value of $q_\rho$ than what the numerical setting introduces. 
In calculating the value of $\chi_{\rm j}$ there is in addition the uncertainty in the value of $\eta \zeta$.
  
\section{Summary}
\label{sec:Summary}

We conducted 3D hydrodynamical simulations to further explore the role of jets in CEJSNs. In our numerical simulations we assumed that a NS companion orbits inside the outer envelope zones of a RSG star at a constant orbital radius of $a=700 R_\odot$ and launches jets perpendicular to the orbital plane $z=0$. We simulated 5 cases varying the power of the jets and the numerical resolution (Table \ref{Table:cases}). Due to the complicated flow structure with the energetic jets that NS companions launch we had to limit our numerical resolution and maximum jets' power. For numerical limitations and for our goal to reveal the role of jets, we did not include the gravity of the NS, nor the self gravity of the envelope, nor the spiralling-in process of the NS inside the RSG. 

We emphasise that we expect that in most CEJSN events the jets are more energetic than the maximum power we could simulate in RUN42L. 

We present the flow structure after two orbital periods of simulation RUN40 in Fig. \ref{fig:RUN40profiles}. This simulation has the most energetic jets that we could study with the high numerical resolution. In Figs. \ref{fig:RUN41vs42} and \ref{fig:RUN41vs42Vel} we present the results for the more energetic simulations RUN41, RUN41L, and RUN42L, but that had lower resolutions (fifth column of Table \ref{Table:cases}). 

From the flow structure we conclude the following (section \ref{sec:flow}).  (1) In most CEJSN events we expect the NS-RSG binary system to experience the GEE before it enters a full CEE. (2) The jets induce downward-flow and upward-flow streams in the RSG envelope. These add to the already existing strong convection in the RSG envelope. The stronger convection will be very efficient in transferring energy in the envelope (e.g., \citealt{Gricheneretal2018,  WilsonNordhaus2020}). Future numerical simulations of CEE with a NS/BH companion should include energy transport by the strong convection.
 
We calculated the average density of a thin spherical shell of radius $a=700 R_\odot$. This value fluctuates with time (Fig. \ref{fig:AverageD}). After the initial large fluctuations relaxed, we averaged the density over a time period that we mark by the horizontal dashed-dotted black lines on Fig. \ref{fig:AverageD}. We then calculated the factor $q_\rho$ by which the jets reduce the average density (equation \ref{eq:ChiJ}), and took this ratio to be the negative jet feedback coefficient $\chi_{\rm j}$. We list these values in the sixth column of Table \ref{Table:cases}. Because we could not simulate high-power jets, in Fig. \ref{fig:jetspower} we connected the results of our 3D simulations with the 1D simulations of \cite{Gricheneretal2021}. This allowed us to estimate the negative jet feedback coefficient that we give in equation (\ref{eq:ChiFinalValue}). This is our main result. 

\cite{Gricheneretal2021} estimated this range to be $0.04 - 0.3$. With the new simulations we narrowed somewhat this range to $\chi_{\rm j} \simeq 0.1-0.2$. However, uncertainties are still large (section \ref{sec:Uncertainties}). 
Future simulations will have to better resolve the NS/BH vicinity, to include the gravity of the NS/BH, to include the self gravity of the envelope, and most of all to include in the simulations the actual accretion rate onto the NS/BH and vary the jets' power accordingly.   

\section*{Acknowledgments}
We thank Aldana Grichenr and Sagiv Shiber for helpful discussions, and an anonymous referee for useful comments. 
This research was supported by the Pazy Foundation.

\section*{Data availability}

The data underlying this article will be shared on reasonable request to the corresponding author. 

\label{lastpage}
\end{document}